%% file: main.tex
\pdfoutput=1
\documentclass[preprint]{sigplanconf}

\renewcommand{\topfraction}{0.99}

\usepackage{amsmath}
\usepackage{afterpage}

\usepackage{subfig}
\usepackage{graphicx}

\usepackage{makeidx}  %
\usepackage[usenames,dvipsnames]{xcolor}
\usepackage[]{hyperref}
\usepackage{fancyvrb}
\usepackage{pdflscape}
\usepackage{listings}
\usepackage{pifont}
\usepackage{enumitem}

\usepackage{hyperref}

\makeatletter
\def\@copyrightspace{\relax}
\makeatother

\begin{document}

\lstset{
  language=C++,
   basicstyle={\ttfamily},
   keywordstyle={\ttfamily\bf},
   numberstyle = {\scriptsize\sffamily},
   commentstyle = {\sffamily\emph},
  columns    = flexible,
  captionpos = b,
  numbers     = left,
  numberblanklines={false},
  escapechar=@,
  morestring=[d]'',
  xleftmargin=1.5em,
  breaklines={false},
  morecomment=[l]{//},
  morecomment=[s][\footnotesize\sffamily\color{red}\emph]{//+}{+//},
  moredelim=**[il][\color{red}]{(r)},
  moredelim=**[il][\color{orange}]{(b)},
  showstringspaces=false,
 numbersep=5pt
}

\title{Hardware extensions to make lazy subscription safe\thanks{Copyright
    \textcopyright{} 2014, Oracle and/or its affiliates.  All rights
    reserved.}}  
\date{}

 \authorinfo{Dave Dice}
            {Oracle Labs}
            {dave.dice@oracle.com}

 \authorinfo{Timothy L. Harris}
            {Oracle Labs}
            {timothy.l.harris@oracle.com}

 \authorinfo{Alex Kogan}
            {Oracle Labs}
            {alex.kogan@oracle.com}

 \authorinfo{Yossi Lev}
            {Oracle Labs}
            {yossi.lev@oracle.com}

 \authorinfo{Mark Moir}
            {Oracle Labs}
            {mark.moir@oracle.com}

\maketitle
\input{abstract}
\input{intro}

\input{lazy}

\input{flaws}

\input{hardware}
\input{conc}

\bibliographystyle{plain}
\bibliography{lazysub}

\end{document}

%% file: abstract.tex
\begin{abstract}
  Transactional Lock Elision (TLE) uses Hardware Transactional Memory
  (HTM) to execute unmodified critical sections concurrently, even if
  they are protected by the same lock.  To ensure correctness, the
  transactions used to execute these critical sections ``subscribe''
  to the lock by reading it and checking that it is available.  A
  recent paper proposed using the tempting ``lazy subscription''
  optimization for a similar technique in a different context, namely
  transactional systems that use a single global lock (SGL) to protect
  all transactional data.  

  We identify several pitfalls that show that lazy subscription
  \emph{is not safe} for TLE because unmodified critical sections
  executing before subscribing to the lock may behave incorrectly in a
  number of subtle ways.  We also show that recently proposed compiler
  support for modifying transaction code to ensure subscription occurs
  before any incorrect behavior could manifest is not sufficient to
  avoid all of the pitfalls we identify.  We further argue that
  extending such compiler support to avoid all pitfalls would add
  substantial complexity and would usually limit the extent to which
  subscription can be deferred, undermining the effectiveness of the
  optimization.

  Hardware extensions suggested in the recent proposal also do not
  address all of the pitfalls we identify.  In this extended version
  of our WTTM 2014 paper, we describe hardware extensions that make
  lazy subscription safe, both for SGL-based transactional systems and
  for TLE, without the need for special compiler support.  We also
  explain how nontransactional loads can be exploited, if available, to
  further enhance the effectiveness of lazy subscription.
\end{abstract}

%% file: intro.tex
\section{Introduction}

Hardware Transactional Memory \cite{HM:93,rock-micro-2009,Intel-SC13} provides
hardware support for atomically executing a section of code, without
requiring programmers to determine how this atomicity is achieved.
Numerous techniques for exploiting HTM to improve the performance and
scalability of concurrent programs have been described in the
literature \cite{rlsapps,SLE,ASPLOS-tech-report-2009,Dragojevic-HTMDynamicCollect:2011,Dice-SimplifyingHTM:SPAA2010}.

The simplest and most readily exploitable of these techniques is
Transactional Lock Elision (TLE) \cite{rlsapps,SLE}, which targets
existing lock-based applications without requiring them to be
restructured and without modifying critical section code.  TLE uses a
hardware transaction to atomically apply the effects of a critical
section without acquiring the lock, thereby allowing other critical
sections protected by the same lock to be similarly executed in
parallel, provided their data accesses do not conflict.

Because hardware transactions may fail due to conflicts or to
limitations of the HTM implementation, some critical sections must
still be executed in the traditional manner (i.e., not in a hardware
transaction) after acquiring the lock.  To ensure that a critical
section executed in a hardware transaction does not observe partial
effects of a critical section executed by another thread that
acquires the lock, the transaction ``subscribes'' to the lock, i.e.,
it reads the lock and confirms that it is available.
Similar techniques can be used to implement a transactional memory
system in which all transactional data is protected by a single global
lock (SGL), and transactions are executed either by acquiring the
lock, or within a hardware transaction that subscribes to the lock.

Subscribing to the lock makes hardware transactions vulnerable to
abort if another thread acquires the lock.  Typically, transactions
subscribe to the lock at the beginning of the critical section and are
thus vulnerable to such abort during the entire execution of the
critical section.  It is therefore tempting to use a \emph{lazy
  subscription} optimization
\cite{Dalessandro:Hybrid-NOrec-ASPLOS-2011}, which delays lock
subscription, in order to reduce the duration of this vulnerability.
Calciu et al.  \cite{irina-transact2014} recently proposed to use this
technique for SGL-based transactional systems.

A simple (but incorrect) way to implement lazy subscription for TLE is
to delay subscription until immediately before committing the
transaction.  This way the implementation affects only library code
and does not require analysis or modification of critical section
code, retaining the key advantage of TLE that makes it the most
promising way to exploit HTM in the near future.

One might reason that this ``lazy subscription'' technique is safe for
TLE on the grounds that the hardware transaction ensures that all of
the memory accesses performed by the critical section, together with
the check that the lock is not held, are performed atomically, and
therefore the effects of committing the transaction are identical from
the perspective of other threads.  Unfortunately, as we show, there
are subtle problems with this reasoning.  In fact, TLE with lazy
subscription is subject to a number of pitfalls that can violate
correctness by changing the application's semantics.

Because SGL-based transaction systems generally entail static analysis
of all code potentially executed within transactions, there is an
opportunity for the compiler to recognize situations in which
transactions will potentially behave incorrectly, and to ensure they
subscribe to the lock before allowing this possibility.  However, the
analysis proposed by Calciu et al. \cite{irina-transact2014} is not
sufficient to avoid all of the pitfalls we identify.  Furthermore, we
argue that it is unlikely to be practical to enhance the static
analysis to make lazy subscription safe while retaining its benefits
because subscription will be required relatively early in all but very
simple cases.

Hardware extensions are briefly described in \cite{irina-transact2014}
that the authors claim would allow these issues to be avoided
entirely.  However, their extensions are not sufficient to avoid all
of the pitfalls we describe.  In this paper, we describe
ways in which HTM implementations could be enhanced to make lazy
subscription safe for both TLE and SGL-based transaction
implementations, without special compiler analysis.  We also explain
how the technique can be even more effective if the extended HTM
implementation supports nontransactional loads.

While we believe that the hardware enhancements we describe 
are practical and implementable, they
will entail nontrivial cost and complexity.  In ongoing work, we are
exploring the value of the lazy subscription optimization to help
assess whether such hardware extensions are likely to be justified; 
preliminary results suggest that it provides significant performance
benefit in at least some cases.
Collectively, our work in this area contributes to
understanding of the problem and potential solutions, and to
consideration of whether the benefits of such optimizations justify
the cost and complexity required by hardware extensions to make them
safe.

%% file: lazy.tex
\section{TLE and lazy subscription}
\label{sec:lazy}

\input{fig-tle}

TLE is typically implemented by modifying lock library code so that the lock
acquire method begins a transaction, checks that the lock is
available, and if so allows the critical section to execute
without acquiring the lock.  This lock ``subscription'' adds the
lock to the transaction's read set, so that the transaction will
abort if the lock is subsequently acquired before it
commits.  If the lock is not available, the transaction is aborted and
the critical section execution attempt is retried, either in another
hardware transaction or by acquiring the lock and executing the
critical section as usual.  The lock release method commits the
transaction if the critical section was executed in a transaction and
releases the lock otherwise.

This arrangement is illustrated in pseudocode on the left side of
Figure~\ref{fig:tle}, where \texttt{use\_TLE} and
\texttt{using\_TLE} abstract away practical details such as whether
and how long to back off before retrying, whether to wait for the lock
to be available before retrying, how many attempts to make using HTM
before giving up and acquiring the lock, supporting nesting, and how
the \texttt{release} method determines whether the \texttt{acquire}
method chose to use TLE.  These issues are not relevant to
correctness, which is our focus here; some of them are 
explored in detail in \cite{ALE-SPAA2014}.

A TLE transaction executed using the simple technique illustrated on
the left of Figure~\ref{fig:tle} has the lock in its read set
throughout the execution of the critical section.  Thus, \emph{any}
critical section that acquires the lock in this entire duration will
cause the transaction to abort.  One might consider it an
advantage to abort such transactions earlier, given that they may
waste less work in this case.  However, this reasoning overlooks the
fact that in many cases the abort is not necessary (for example
because the critical sections executing in the transaction and with
the lock held do not conflict), so avoiding it is preferable.

Simple lazy subscription, illustrated on the right side of
Figure~\ref{fig:tle}, moves subscription from the \texttt{acquire}
method to the \texttt{release} method, allowing the transaction to
execute the entire (unmodified) critical section without subscribing,
with the understanding that it would do so before committing.

Unfortunately, if a critical section executed in a transaction
observes values in memory that it could not observe if all critical
sections were executed while holding the lock, then it may behave
differently than is intended by the programmer who wrote the critical
section code.

One might argue that this is not a problem, as follows:
The transaction will try to commit only after subscribing to the lock
and observing that the lock is available, implying that its read set
has a consistent view of memory.  Therefore, if the transaction saw an
inconsistent view of memory, then the normal HTM mechanisms will cause
it to abort.
This is the essence of the ``intuititive'' correctness
argument in \cite{irina-transact2014}.  But this incorrectly assumes
that the transaction will eventually execute the correct subscription
code and observe the correct lock state before attempting to commit.
If this is not the case, then the transaction may erroneously commit, 
with unpredictable effects.  We discuss a
number of ways in which the transaction may fail to correctly
subscribe to the lock in the next section.

%% file: fig-tle.tex
\begin{figure}[t]
\begin{minipage}{0.475\linewidth}
\begin{lstlisting}
acquire(lock L) {
retry:
  while (use_TLE(L)) {
    txbegin retry;
    if (isLocked(L)) 
      txabort;
    return;
  }
  < acquire lock L >
} 

release(lock L) {
  if (using_TLE(L)) 
    txcommit;
  else
    < release lock L >
}
\end{lstlisting}
\end{minipage}
\ \ 
\begin{minipage}{0.475\linewidth}
\begin{lstlisting}
acquire(lock L) {
retry:
  while (use_TLE(L)) {
    txbegin retry;
    return;
  }
  < acquire lock L >
} 

release(lock L) {
  if (using_TLE(L)) {
    if (isLocked(L)) 
      txabort;
    txcommit;
  } else
    < release lock L >
}
\end{lstlisting}
\end{minipage}
\vspace*{-.2in}\caption{Pseudocode showing basic TLE (left) and lazy
  subscription version (right).  The \texttt{txbegin} instruction
  specifies a label to which control branches if the transaction
  aborts for any reason.  The \texttt{use\_TLE} method represents a
  policy decision about whether to use TLE; \texttt{using\_TLE}
  returns the value most recently received by the thread from
  \texttt{use\_TLE}.  These methods take an argument identifying the lock to
  enable support for general locking patterns; this is not needed if locking
  is assumed to be properly nested.}
\label{fig:tle}
\end{figure}

%% file: flaws.tex
\section{Pitfalls of lazy subscription}
\label{sec:causes}

Lazy subscription can cause a transaction to deviate from behavior
allowed by the original program in a variety of ways.  Some of these
behaviors are benign, because the transaction aborts and therefore its
effects are not observed by other threads.  In particular, most HTM
implementations ensure that, if a transaction executes code---such as
divide-by-zero---that would ordinarily cause the program to crash, it
simply aborts.  However, below we explain a number of ways in which
a transactions that deviate from the original program's behavior 
can commit successfully, resulting in observably
incorrect behavior\vspace{.1in}.

\noindent\textbf{Observing inconsistent state} If a thread executes a 
critical section without acquiring or subscribing to the lock, this
can result in the thread's registers containing values that could
never occur in an execution of the original program.  This is
illustrated by the example in
Figure~\ref{fig:indirect-branch-example}, in which a shared variable
\texttt{next\_method} indicates the method to perform next time 
\texttt{apply\_next} is invoked.  If the critical section is
executed in a transaction with lazy subscription, at
line~\ref{getnextmethod} it may observe the value of
\texttt{next\_method} as 2 because another thread that is executing
the critical section while holding the lock is just about to reset
\texttt{next\_method} to zero (at line~\ref{resettozero}).  The use of
the lock in the original program ensures that no thread ever reads 2
from \texttt{next\_method}\vspace{.1in}.  

Below we describe a number of ways in which such inconsistent state
can lead to observably incorrect behavior\vspace{.1in}.

\noindent\textbf{Indirect branch} 
Continuing the example above, after a transaction reads 2 from
\texttt{next\_method}, it reads the value stored immediately
\emph{after} the \texttt{method\_table} array and treats it as a
function pointer, invoking the ``code'' at that address.  Because this
address may point to any code or data, the result of executing code
stored at the address is unpredictable.  In particular, it
might 
commit the
transaction, without ever subscribing to the lock\vspace{.1in}.

\input{fig-indirect-branch-example}

This example shows that a thread executing a critical section in a
transaction that has not yet subscribed to the lock can observe values
in memory that it could never observe in any execution of the original
program and that it can commit nonetheless, resulting in observably incorrect
behavior.  While this is sufficient to conclude that lazy subscription
cannot be blindly used for TLE with unmodified critical section code,
it is important to understand that there are many other ways in which
reading inconsistent values from memory can indirectly result in
incorrect behavior, as described below\vspace{.1in}.

\noindent\textbf{Propagating inconsistent state} Once a thread's
registers are in a state not allowed in the original program, this
inconsistency can propagate through the thread's state in numerous
ways, resulting in differences from behavior that could be observed in
an execution of the original program:
\begin{itemize} 
\item Inconsistent values may propagate between registers via
  arithmetic operations, register moves, etc.
\item Inconsistent values in registers may propagate to memory written
  by the transaction explicitly or implicitly (e.g., arguments to
  method calls, register spills).
\item Inconsistent register values may be used as addresses for stores
  to memory, resulting in locations being written that would not be
  written by the transaction in an execution of the original program.
\item Inconsistent values written to memory or to inconsistent
  locations may propagate back to registers via loads, either
  explicitly or implicitly.
\item Conditional control flow may differ.
\end{itemize}

\noindent These effects are benign if the transaction aborts, but they can lead
to the transaction committing without subscribing to the lock in a
number of ways, some of which are discussed below\vspace{.1in}.

\noindent\textbf{Conditional code that commits the transaction} If a
condition in a transaction executing before subscribing to the lock
evaluates differently because of an inconsistent value in a register,
then a code path may be executed that would not be executed by the
original program.  Because we assume arbitrary, unmodified critical
section code, we cannot rule out the possibility that this code could
commit the transaction without subscribing to the lock\vspace{.1in}.

\noindent\textbf{Lock scribbling} A memory write that uses an
inconsistent register for its target address may overwrite the lock
protecting the critical section with a value that makes it appear to
be available.  In this scenario, even if the correct lock subscription
code is executed and subscribes to the correct lock, it may
incorrectly conclude that the lock is available and commit the
transaction\vspace{.1in}.

\noindent\textbf{Subscribing to the wrong ``lock''} If the address of
the lock protecting the critical section is stored in a register or
memory location that is inconsistent, then even if the correct
subscription code is executed, the transaction may incorrectly
conclude that the lock is available and commit\vspace{.1in}.

\noindent\textbf{Self modifying code} Similar to lock scribbling, if a
transaction that has observed inconsistent state writes incorrect
values to memory, or writes to an incorrect address, the transaction
could execute code that it has itself incorrectly written.  Again,
this could result in committing the transaction without subscribing to
the lock\vspace{.1in}.

\noindent\textbf{Corrupted return address}  Finally, we present 
one more concrete example showing how an inconsistent value read from
memory can propagate to cause the transaction to commit without
subscribing to the lock.  In this example, similar to the indirect
branch example above, a transaction using late subscription reads a
value from memory that it could never read in the original program.
This time, it uses this value as an index into a
\emph{stack-allocated} array and writes to memory at the indexed
location.  In this case, if the inconsistent value is not a valid
index into the array, the target location may happen
to be the stack location containing the function's return address, and
the value written may happen to be the address of the instruction that
commits the transaction.  When the function returns, it will execute
the instruction to commit the transaction without attempting to
subscribe to the lock.

\subsection{Avoiding the pitfalls via compiler support}

TLE is the most promising way to exploit HTM in the short term because
it can be applied to unmodified critical sections, with no special
compiler support.  (Note that modifying critical sections may be
required in order to achieve the best performance, but not to ensure
correctness.)  As explained above, lazy subscription cannot be applied
to TLE without sacrificing this important property.

For the context of SGL-based transactional systems, compiler support
for analyzing code to be executed in transactions is typically
required anyway, so there is an opportunity for the compiler to
analyze and modify such code in order to make lazy subscription safe.
Indeed, Calciu et al.  \cite{irina-transact2014} proposed that the
compiler ensure that transactions subscribe to the lock before
executing an indirect branch in order to avoid the indirect branch
pitfall described above.  (We note, however, that they suggested this
only for the case in which the transaction had already written to
memory; the indirect branch example above shows that this is not
sufficient, as it does not write to memory
before executing the indirect branch.)

Presumably they also assumed that the compiler would conservatively
disallow the use of instructions that would commit the transaction
within any code that could \emph{potentially} be executed within a
transaction.  This would avoid the ``conditional code that commits the
transaction'' pitfall.

However, Calciu et al. did not identify 
the remaining pitfalls described above, nor did they propose
any mechanisms that would avoid them.  Given the diverse range of
ways in which a transaction may commit incorrectly, we would argue
that any static analysis that is sufficient to ensure correctness
would entail significantly more complexity than is suggested in
\cite{irina-transact2014}.  

The complexity required by such static analysis may be mitigated to
some degree by conservatively subscribing to the lock to avoid the
need to precisely determine whether the transaction may violate
correctness in various cases.  However, this reduces the effectiveness of
the lazy subscription optimization.  

Given the numerous ways in which inconsistency can propagate and
manifest, even maximally precise analysis will likely often require
relatively early subscription.  For example, the corrupted return
address pitfall suggests that subscription is necessary before the
first time a transaction returns from a function call after reading a
potentially-inconsistent value from memory and subsequently performing
a write, even to its own stack.  Applying this rule precisely 
requires analysis that ensures any record of whether the transaction
has previously read from memory is accurate.

Similarly, avoiding the ``subscribing to the wrong lock'' pitfall
requires the transaction to ensure that its notion of which lock it is
eliding is not corrupted by propagating inconsistent data.  Avoiding
``lock scribbling'' requires not only a reliable record of the lock's
address, but also knowledge of the structure of the lock, unless the
compiler is so conservative that it does not allow \emph{any} writes to
memory based on a potentially-inconsistent address register.

Clearly at least some safe deferral of lock subscription is
\emph{possible} with sufficiently precise or conservative analysis.
However, we believe the complexity required to make lazy
subscription safe using software techniques alone is unlikely to be
worthwhile for the degree to which subscription can be deferred in
practice.

Finally, we note that hardware extensions briefly described in
\cite{irina-transact2014} are not sufficient to avoid all of the
pitfalls described above.  In particular, although the proposed
extensions ensure that the correct lock is subscribed to before a
transaction commits, there is no mechanism proposed to avoid the
``lock scribbling'' pitfall.

%% file: fig-indirect-branch-example.tex
\begin{figure}[t]
\begin{minipage}{0.95\linewidth}
\begin{lstlisting}
void (*method_table[2])() = {method1, method2};

int next_method = 0;

lock L;

void apply_next() {
  acquire(L);
  (*method_table[next_method])();   @\label{getnextmethod}@
  if (++next_method > 2)
    next_method = 0;   @\label{resettozero}@
  release(L);
}
\end{lstlisting}
\end{minipage}
\vspace*{-.2in}\caption{An example in which an indirect branch
  executed within a transaction has an unrpedictable target.}
\label{fig:indirect-branch-example}
\end{figure}

%% file: hardware.tex
\section{Making lazy subscription safe and effective}
\label{sec:proposals}

The essence of all of the hardware approaches we describe for supporting lazy
subscription is to ensure that the lock and the method for subscribing
to it are identified \emph{before} beginning transactional execution
of a critical section, and to ensure that the transaction correctly
subscribes to the identified lock using the identified method before
committing, regardless of what code the transaction executes.  
(For generality, we note that in fact this and other
information discussed below only needs to be recorded before any actions that
could potentially corrupt the information being recorded.  However, because recording
this information does not make the transaction more vulnerable to abort, it is unlikely
to be worthwhile to complicate an implementation in order to delay this reording.)
We begin with a simple approach and then present more complex approaches
that address its limitations.  

\subsection{A simple but inflexible approach}

First, it is preferable that transactions are limited to execute only
for a bounded number of instructions or cycles. This avoids the
possibility that a critical section that is executed with lazy
subscription goes into an infinite loop due to observing transient
data. Without this restriction, another solution would be needed to
avoid this possibility, such as requiring transactions to subscribe to
a special variable that is periodically modified. Most
or all existing HTM implementations already limit transaction length.

In our simple approach, we next add a special register,
called the \emph{lock address register} (LAR), which is set to the
address of the lock before beginning transactional execution of a
critical section.  Any attempt to modify the contents of the LAR
during transactional execution causes the transaction to abort.  Any
attempt to commit an outermost hardware transaction (i.e., one that is
not nested within another hardware transaction) causes the location
identified by the LAR to be read transactionally and compared to zero;
if the comparison fails, then the transaction is aborted.
Furthermore, any attempt by the
thread executing the transaction to write to
the memory location whose address is stored in the LAR causes the
transaction to abort.  This approach is simple to implement, but 
suffers from several severe limitations.

\subsection{Limitations of the simple approach}

The simple approach desribed above supports only locks that represent
the ``available'' state by storing zero at the address used to
identify the lock.  Some other locks could be supported by the
addition of another register that is similarly set before the
transaction and not modifiable during it, which would store a bitmask
to use to check lock availability; for example, this would support
seqlocks \cite{linux-seqlock,lameter-seqlock-2005}, 
which use only a single bit to represent lock
availability, while storing additional information in other bits (the
sequence number in the case of seqlocks). 

Nonetheless, many other important lock types are not supported so
easily.  For example, ticket locks
\cite{Dice:partitioned-ticket-lock,MCS:91} require two values to
be compared to test lock availability, local-spin locks such as CLH
\cite{Craig-lock,MLH-lock} require a pointer to be dereferenced and 
the pointed-to value tested for availability, etc.  

Although a conservative approximation of lock availability suffices to
preserve correctness, it may reduce or eliminate the benefit of TLE.
For example, some lock types \cite{Agesen-metalocks,Bacon-thinlocks} represent the ``available''
state as zero until the lock experiences contention, at which point it
is ``inflated''\negthinspace, requiring a pointer to be dereferenced
to accurately determine lock availability.  Simple schemes like the
one described above would thereafter always determine that the lock is
not available, thus permanently eliminating the benefit of TLE.

In principle, arbitrarily complex subscription methods could be baked
into hardware, so that they could not be modified by critical section
code that has observed transient data.  However, it is clearly
preferable to be able to express subscription methods in software, as
discussed further below.

The simple approach is also limited in that it does not fully support
lazy subscription for nested critical sections: if the LAR has
already been set to ensure lazy subscription of the lock for one
critical section, then it would not be possible to achieve lazy
subscription of a nested critical section protected by a different
lock.  It is not difficult to extend the ideas described above to
support a fixed number of nesting levels by allowing multiple LARs
and, if applicable, associated bitmasks and/or subscription methods.
Alternatively, protected memory area(s)---specified by base and size
registers that are protected as described above---could allow a set of addresses
and associated bitmasks and/or subscription methods to be stored; any
attempt to reduce the size of the protected memory area, or to modify
locations in it or locations identified by it would cause transaction
abort.  

We note that it is possible that, due to observing transient data, a
nested critical section may be configured to use the wrong lock
subscription method or the wrong lock.  This is not a problem,
however, because this can happen only as a result of observing
transient data protected by the lock associated with an enclosing
critical section.  This implies that at least one enclosing critical
section was correctly configured to subscribe to the correct lock
before the transient data was observed.  The nested transaction is
allowed to commit only if \emph{all} of the nested critical sections
successfully subscribe to their locks before committing, and this is
guaranteed not to be the case for the (at least one) lock that is
correctly subscribed.

\subsection{More flexible approaches}

To support arbitrary lock types, we add anther register, which is
managed and protected against corruption similarly to the LAR
discussed above; this \emph{subscription code address
  register} (SCAR) identifies the code for subscribing to the lock
identified by the LAR. (If nesting of different lock types that
require different subscription code is desired, similar techniques as
described above for managing nested locks can be used to record their 
addresses.)

To ensure correct subscription, we must ensure that the critical
section cannot overwrite the subscription code and that it cannot
modify data that the subscription code reads; the latter is necessary to avoid the lock
scribbling pitfall. On the surface, this seems challenging because the
hardware cannot predict which code will be executed when the function
identifued by the SCAR is invoked, nor what data it will access.

An important insight into these issues is that it is not necessary to
abort a transaction as soon as it \emph{writes to} the lock contents
or the subscription code.  We must ensure only that it does not commit
successfully without correctly subscribing to the lock.  Thus,
attempts to overwrite lock data or subscription code need not be
detected until the subscription method attempts to \emph{execute} the
modified code or to \emph{read} the modified lock data.

Therefore, a central aspect of our approach to supporting flexible,
software-defined lock subscription is to enter a mode immediately
before starting to execute the subscription code in which, if the
transaction attempts to execute code or to read data that is in the
transaction's write set, the transaction aborts and does not take
effect.  Because HTM implementations must generally detect cases in
which a transaction reads data it has written, supporting this
behavior does not add significant additional complexity to an HTM
design.

As a side note, while transactions could conceivably be used to
simplify techniques based on self-modifying code by ensuring sets of
changes take effect atomically, we believe that the benefits (if any)
of being able to modify and execute code within the same transaction
are outweighed by the likelihood of such questionable practices
resulting in incorrect behavior.  Therefore, it may make sense to
prevent transactions from executing code they have modified,
independent of the lazy subscription technique.  In contrast,
aborting a transaction because it reads data that it has written
clearly does not make sense in general, so this behavior should be
limited to the execution of lazy subscription code.

We note a potential disadvantage, namely that a transaction might be
caused to abort unnecessarily if it modifies data that is near the
lock, but not actually part of the lock.  This could happen, for
example, if the lock is co-located with data it protects, for example
in the same cache line (if this is the granularity at which a
transaction's write set is tracked).

This does not compromise correctness; it is only a performance issue,
albeit a potentially significant one.  The issue could be mitigated,
at the expense of additional hardware cost and complexity, by
maintaining state for each cache line modified by a transaction that
records at finer granularity---per word, for example---which parts of
the cache line have been modified by the transaction.  Doing so would
allow the subscription method to avoid aborting a transaction that has
modified data in the same cache line as some data read by the
subscription method, even though it has not modified any data actually
read by the subscription method.

A similar approach was suggested by Tabba et
al. \cite{tabba-transact-2009-value-prediction} for the purpose of
avoiding unnecessary transaction aborts due to false sharing.

\subsection{Further extensions}

The purpose of the lazy subscription technique is to reduce the
window in which a transactionally-executed critical section is
vulnerable to abort due to the lock being held or acquired.  We
observe that, if a transaction determines that the lock is held when
it performs this subscription, it is immediately doomed to abort and
retry.  This could be mitigated by techniques that allow a transaction 
to wait for a variable to change value, without aborting.

For example, if the HTM supports nontransactional loads, then in some cases it is
possible to use them to wait for the lock to become available before
subscribing to the lock.  Such waiting does not compromise the
correctness of the subscription, because the lock would ultimately be
subscribed to transactionally before committing the transaction.  As a
simple example, if the lock is a single word representing
``available'' and ``locked'' states, the subscription method would
repeatedly read the word using nontransactional loads until the lock
state is ``available''\negthinspace, and would then load the lock word
transactionally, and confirm that it is available before committing
the transaction.

The effectiveness of such approaches of course depends on the
availability of hardware features on the relevant platform to support
waiting until a variable's value changes without aborting a
transaction.  We recommend that designers of future HTM features
consider whether their design would effectively support such
techniques.

Independent of the lazy subscription technique, our discussions of
the use of nontransactional memory operations within hardware
transactions raise an important observation.  If an HTM feature
supports nontransactional \emph{stores} (or any kind of side effect
that may affect program semantics when executed in a transaction that
aborts), then care must be taken not to use such instructions within
critical sections to be used with TLE.  The reason is that, if an
attempt to execute such a critical section in a hardware transaction
via TLE fails, then the store may take effect even though the critical
section has not been executed yet.  This could result in program
behavior that would not be possible if critical sections were always
executed while holding the lock, breaking the TLE technique.

While it may seem that such nontransactional store instructions would
generally be used only in code that is intended to be explicitly used
in transactions, it is not beyond the realm of possibility that some
code intended for use in hardware transactions might also be called in
critical sections protected by a lock, in which case using TLE to
elide such critical sections would change program semantics.

This observation may motivate support for a transaction execution mode
that insists that all store instructions---even nontransactional ones---are
executed transactionally; this mode would be used for TLE.  In the
absence of such protection, any nontransactional store feature needs
to be used with care to ensure this scenario does not occur.

%% file: conc.tex
\section{Concluding remarks}
\label{sec:conc}

We have discussed a number of ways in which the ``lazy subscription'' optimization for
Transactional Lock Elision (TLE)---in which lock subscription is
delayed until the end of transactional critical section execution in
order to reduce the transaction's window of vulnerability to
abort---is not safe in general with existing hardware transactional memory (HTM)
features.  A transaction may observe inconsistent data
if it does not subscribe to the lock early, and as a result may fail
to correctly subscribe to the lock before committing.

Dalessandro et al. \cite{Dalessandro:Hybrid-NOrec-ASPLOS-2011} first
proposed lazy subscription and pointed out that a hardware transaction
must ensure its reads are consistent before executing any instructions
that may be dangerous if executed based on inconsistent reads.  The
Reduced NOrec algorithm of Matveev and Shavit
\cite{Matveev:Reduced-NOrec-SPAA-2013} recognizes the same issue, and
explicitly separates out cases that are not compatible with lazy subscription
in order to allow lazy subscription for the other (hopefully common) cases.
Specifically, it 
introduces a ``slow path'' that applies the effects of software
transactions using HTM, allowing ``fast-path'' transactions to use
lazy subscription with respect to these transactions.  Nonetheless, in
order to avoid pitfalls such as those described in our paper, fast-path
transactions must subscribe early to a global lock used to protect
``slow-slow-path'' transactions that cannot be committed using HTM;
such transactions are executed in software, and
thus may expose partial effects to hardware transactions that have not subscribed to the lock.

Compiler support suggested recently \cite{irina-transact2014} for avoiding
such issues in SGL-based transaction systems
is not sufficient to ensure
correctness.  We argue that the complexity required to address
these issues via static analysis is unlikely to be worthwhile.  Precise analysis of when
subscription can be deferred is complex and is likely to result in
relatively early subscription in most cases; conservative analysis
to mitigate such complexity will only exacerbate the
problem, largely eliminating any benefit from lazy subscription.

Without detailed analysis of the compiled code for benchmarks used to
evaluate the benefits of lazy subscription, it is difficult to assess
how meaningful their results are.  However, in our ongoing work, we
are experimenting with lazy subscription in carefully controlled
benchmarks for which we are confident lazy subscription does not
compromise correctness.  Our preliminary results convince us that lazy
subscription is worth pursuing further, as it does yield significant
performance benefits without compromising correctness in at least some
cases.  However, as we have argued, there are numerous pitfalls
associated with lazy subscription, so manual confirmation of its
safety in specific cases is likely to be error prone.

In this paper, we have also described hardware extensions that
eliminate these issues entirely in hardware, allowing lazy
subscription to be safely used with TLE and SGL-based transaction
systems with no special compiler support or manual analysis.  While we
believe these changes are likely to add only modest cost and
complexity to an HTM design, such extensions undoubtedly have a cost.
Thus, it remains to be seen whether this cost will be justified by the
benefits of enabling the use of lazy subscription.